# Comment on a suggested Kochen-Specker test

R.L. Schafir[1]

CISM, London Metropolitan University, London EC3N 1JY, U.K.

## Abstract

A suggestion for an observational test of the difference between quantum mechanics and noncontextual hidden variables theories requires the measurement of a product of two commuting observables without measuring either observable separately. A proposal has been made for doing this, but it is shown to be problematic.

Cabello and García-Alcaine [1] have derived some different predictions of quantum mechanics and noncontextual hidden variable theories, but to test their predictions experimentally would require the measurement of the product of two commuting observables without measuring either observable separately. Motivated by this, Simon *et al* [2] have proposed a method of making such a measurement, but doubts will be raised in this Comment about their suggestion.

The idea of Cabello and García-Alcaine, as simplified by Simon *et al*, is as follows. If $Z_1, Z_2, X_1, X_2$ are the the z-direction and x-direction spins of two entangled particles, with their eigenvalues taken to be $\pm 1$, then though the individual products of z and x direction spins do not commute for the same particle:

$$[Z_1, X_1] \neq 0, \;\; [Z_2, X_2] \neq 0 \tag{1}$$

the pair of product observables $Z_1 Z_2$ and $X_1 X_2$ commutes:

$$[Z_1 Z_2, X_1 X_2] = 0 \tag{2}$$

This accords with our knowledge that although we cannot attribute values to the x and z-direction spins for the same particle simultaneously we can attribute to the 2-particle system the results "same" (product eigenvalue 1) or "opposite" (eigenvalue $-1$) in every direction simultaneously. For instance if the particles are in the singlet state then both $Z_1 Z_2$ and $X_1 X_2$ have the value $-1$.

The products $Z_1 X_2$ and $X_1 Z_2$ also commute:

$$[Z_1 X_2, X_1 Z_2] = 0 \tag{3}$$

and it is in different predictions for their values under certain circumstances that there are differences between quantum mechanics and noncontextual hidden variables. It can be shown [1,2] that if the 2-particle system is in the (1, 1) eigenstate of $(Z_1 Z_2, X_1 X_2)$, i.e. their spins are the same for both directions, then in quantum mechanics a measurement of $(Z_1 X_2, X_1 Z_2)$ projects the system into either the (1, $-1$) or ($-1$, 1) eigenstates of $(Z_1 X_2, X_1 Z_2)$. But in noncontextual hidden variables, using the

---

[1] e-mail: roger.schafir@londonmet.ac.uk

assumption [3] that, as in quantum mechanics, for any function $f$ of commuting observables $A,B$:

$$val(f(A,B)) = f(val\ A, val\ B)) \qquad (4)$$

then, since $Z_1, Z_2, X_1, X_2$ all have separate values, which happen to be such that

$$val(Z_1Z_2) = val(X_1X_2) = 1 \qquad (5)$$

it follows that at the same time as (5) holds, the following also holds:

$$val(Z_1X_2) = val(X_1Z_2) = \pm 1 \qquad (6)$$

Therefore, using the further assumption that noncontextual hidden variables shares with quantum mechanics the property that for any values possessed by observables, if there is then a measurement of those same observables the same values are repeated, [2] if $(Z_1X_2, X_1Z_2)$ are then measured, the results

$$val(Z_1X_2, X_1Z_2) = (1,\ 1)\ \text{or}\ (-1, -1) \qquad (7)$$

will be obtained.

But how do we measure a product like $Z_1Z_2$ without measuring $Z_1$ and $Z_2$ separately? To suggest a method, Simon *et al* turn away from entangled states to a situation of two commuting attributes of a single particle, a path and a spin, but for which the above reasoning continues to hold. Taking $Z_1$ now as the path observable, with 1 indicating an "up" path and $-1$ a "down" path, but $Z_2$ still as spin $\pm 1$, the method can be shown in two stages.

Each particle enters an apparatus from either an up path u ($Z_1 = 1$), or a down path d ($Z_1 = -1$), or a superposition of the two, and there are Stern-Gerlach analysers, SG1 and SG2, for each path, each with two outlets for whether the spin is $+1$ or $-1$, so that a detector in each outlet would reveal not only the path but also the spin.

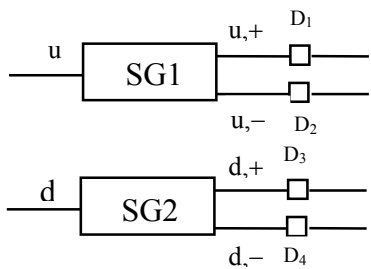

Fig. 1

In the second stage, dispense with the detectors in the outlets of the Stern-Gerlach analysers, and direct the (u, +) and the (d, −) outlets, i.e. the paths which give the value 1 to the product $Z_1Z_2$, into one beam combiner,[3] and the (u, −) and (d, +) outlets into another, by some evolution rule such as

$$\tfrac{1}{\sqrt{2}}\{|u,+\rangle+|d,-\rangle\} \to |\text{BC1outlet}\rangle, \quad \tfrac{1}{\sqrt{2}}\{|u,-\rangle+|d,+\rangle\} \to |\text{BC2outlet}\rangle \qquad (8)$$

and now place detectors after the outlets of the beam combiners.

---

[2] All assumptions about noncontextual hidden variables should be made explicit because of the extreme likelihood that any experiment will verify the predictions of quantum mechanics, so the claim that noncontextual hidden variables would have given a different result will remain a theoretical statement.

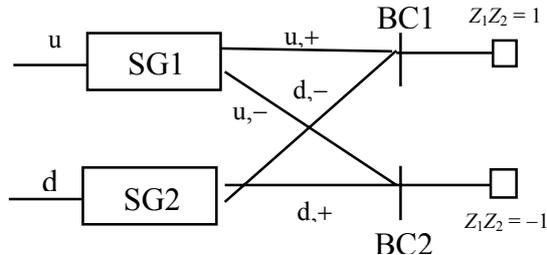

Fig. 2

Then, the argument goes, if there is a detection after BC1, then the particle was in either the $|u,+\rangle$ or $|d,-\rangle$ state (though we don't know which, because of the erasing effect of the combiner), and since in either case the value of the product $Z_1Z_2$ was 1, we can attribute this unique value to $Z_1Z_2$ before the combiner, and hence after the combiner as well, since the value is unchanged by evolution through the combiner. Equally if there is a detection after BC2 then the particle was in $|u,-\rangle$ or $|d,+\rangle$, and so the value of the product $Z_1Z_2$, both before and after the combiner, is −1.

This depends on deducing from the result of a measurement that an observable possessed a certain value *before* that measurement, but this is something which is not in accordance with the principles of orthodox quantum mechanics. This is both in itself and because it can lead to an attribution of simultaneous values for non-commuting observables. A simple case of this is given by the following example, which featured in discussions of the consistent-histories interpretation (see Ref [4] and the further references therein), but is here applied to orthodox quantum mechanics.

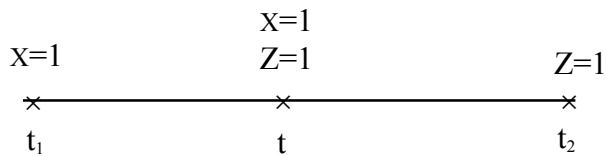

Fig 3

Suppose a spin-½ particle has its x-direction spin measured at time $t_1$, and gives the value (say) +1, and then has its z-direction spin measured at a later time $t_2$, and again gives (say) +1. Suppose (a contrafactual assumption) there had been a second measurement of spin-x at an intermediate time t. Then clearly the result would have to have been +1, while spin-z would have had no value at all (in orthodox quantum mechanics). But equally we could suppose that spin-z had been measured at t, in which case it might appear that the z-result would have had been +1 (so that this value would be repeated at $t_2$), while spin-x would then have had no value. This is a contradiction and seems to require us to abandon the assumption that non-commuting observables have no simultaneous values, or at least weaken the assumption to allow contrafactual attributions of simultaneous values.

---

[3] This is a further simplification of the scheme presented by Simon *et al.*

But this reasoning assumes that the z-result at $t_2$ would have been the same whether or not there was an earlier measurement of z-spin. This is not necessarily the case, because if, contrary to fact, z-spin had first been measured at t instead of at $t_2$, the result at t might have been −1, and this result would have been repeated at $t_2$, instead of obtaining +1 at $t_2$.

Similarly in the present case, the deduction of a value for $Z_1Z_2$ depends on the assumption that the result of a measurement after the particle has been through the beam combiners would have been the same whether or not there had also been a measurement before the beam combiners. If a particle has been discovered in the outlet of BC1, for example, it seems to follow that if detectors had been in the paths before the beam combiners, one of those in the $|u,+\rangle$ or $|d,-\rangle$ paths would have fired. (This is the meaning of the statement that the particle must have been in one of the paths leading to BC1.) But if, contrary to fact, there had been detectors in the paths before the combiners, one of those in the $|u,-\rangle$ or $|d,+\rangle$ paths might have fired, and if so the detector after BC2 would then have fired, instead of the detector after BC1.

What is our present day reason for attributing values to $Z_1Z_2$ and $X_1X_2$ at the same time, in circumstances such as the singlet state, without attributing values to the individual observables themselves? It is because, although only one pair of observables for each particle may be actually measured, and the other pair are contrafactual, the attribution of values is made to them with probability 1. This need not follow only from theory, for in purely experimental terms we could imagine a long series of observations of the results of spin measurements in different directions, from which it is concluded that the results for each particle are always opposite whatever the direction; and on these grounds we attribute the result −1 to the product for every choice of direction simultaneously. But of course it would be an advance to have a direct measurement of the product without a measurement of the individual observables, and even better to have a method of measuring $Z_1Z_2$ and $X_1X_2$ simultaneously. But no method for this appears yet to have been found.

## References

[1]    A. Cabello and G. García-Alcaine, Phys. Rev. Lett. **80**, 1797 (1998).

[2]    C. Simon, M. Zukowski, H. Weinfurter and A. Zeilinger, Phys. Rev. Lett. **85**, 1783 (2000); quant-ph/0009074.

[3]    N. D. Mermin, Rev. Mod. Phys. **65**, 803 (1993).

[4]    R. Omnes, J. Stat. Phys. **62**, 841 (1991), §6.